# Impact-based Analysis Method (IBAM) to evaluate the primary stability of the pedicle screw: a proof of concept

Julie Gabriel[a,b], Yvan Petit[a,d], Giuseppe Rosi[c], Éric Wagnac[a,d], Michel Dagher[e,f], Arnaud Dubory[e,f], Charles-Henri Flouzat-Lachaniette[e,f], , Guillaume Haïat[b]

[a] Department of Mechanical Engineering, École de technologie supérieure, 1100 Notre-Dame Street West, Montréal, Québec H3C 1K3, Canada

[b] CNRS, Laboratoire de Modélisation et de Simulation Multi-Échelle, MSME UMR CNRS 8208, 61 Avenue du Général de Gaulle, Créteil Codex 94010, France

[C] Univ. Paris Est Créteil, Univ. Gustave Eiffel, CNRS, UMR 8208, MSME, F-94010 Créteil, France

[d] Research center, CIUSSS du Nord-de-l'île-de-Montréal, 5400 Gouin blvd, Montréal H4J 1C5, Québec, Canada

[e] INSERM U955, IMRB Université Paris-Est, 51 avenue du Maréchal de Lattre de Tassigny, 94000, Créteil, France

[f] Service of Orthopedic and Traumatological Surgery, Hôpital Henri Mondor AP-HP, CHU Paris 12, Université Paris Est, 51 avenue du Maréchal de Lattre de Tassigny, 94000 Créteil, France

## Abstract

The long-term success of spinal instrumentation depends on the initial stability of pedicle screws (PS), yet current intraoperative assessments rely on surgeon subjectivity. This study introduces a quantitative, non-invasive approach — the Impact-Based Analysis Method (IBAM) — to support intraoperative decision-making. IBAM measures screw stability using low-amplitude impacts delivered with an instrumented hammer on the surgical screwdriver after screw insertion. An indicator, defined as the average time difference between the two earliest local maxima in the force signal, was analyzed under varying stability conditions in artificial bone samples. Experimental factors included bone density (ρ), insertion depth ($D$), and pilot hole diameter ($\phi$). The IBAM indicator was compared with the PS pull-out force for each configuration. Although the main effects showed opposite trends for the pull-out force and the IBAM indicator, their relative magnitudes were similar ($\rho$ ~35%, $D$ ~25%, $\phi$ ~15%). The relationship between pull-out force and the IBAM indicator followed a power-law model (adjusted $R^2 = 0.85$) with a negative exponent. As screw stability increased, pull-out force rose while the IBAM indicator decreased, demonstrating complementary sensitivity to bone-implant interface changes. While results are promising, the simplified experimental design limited the ability to fully capture bone-implant interaction complexity. Future studies should aim to validate IBAM in vivo to confirm its clinical applicability.

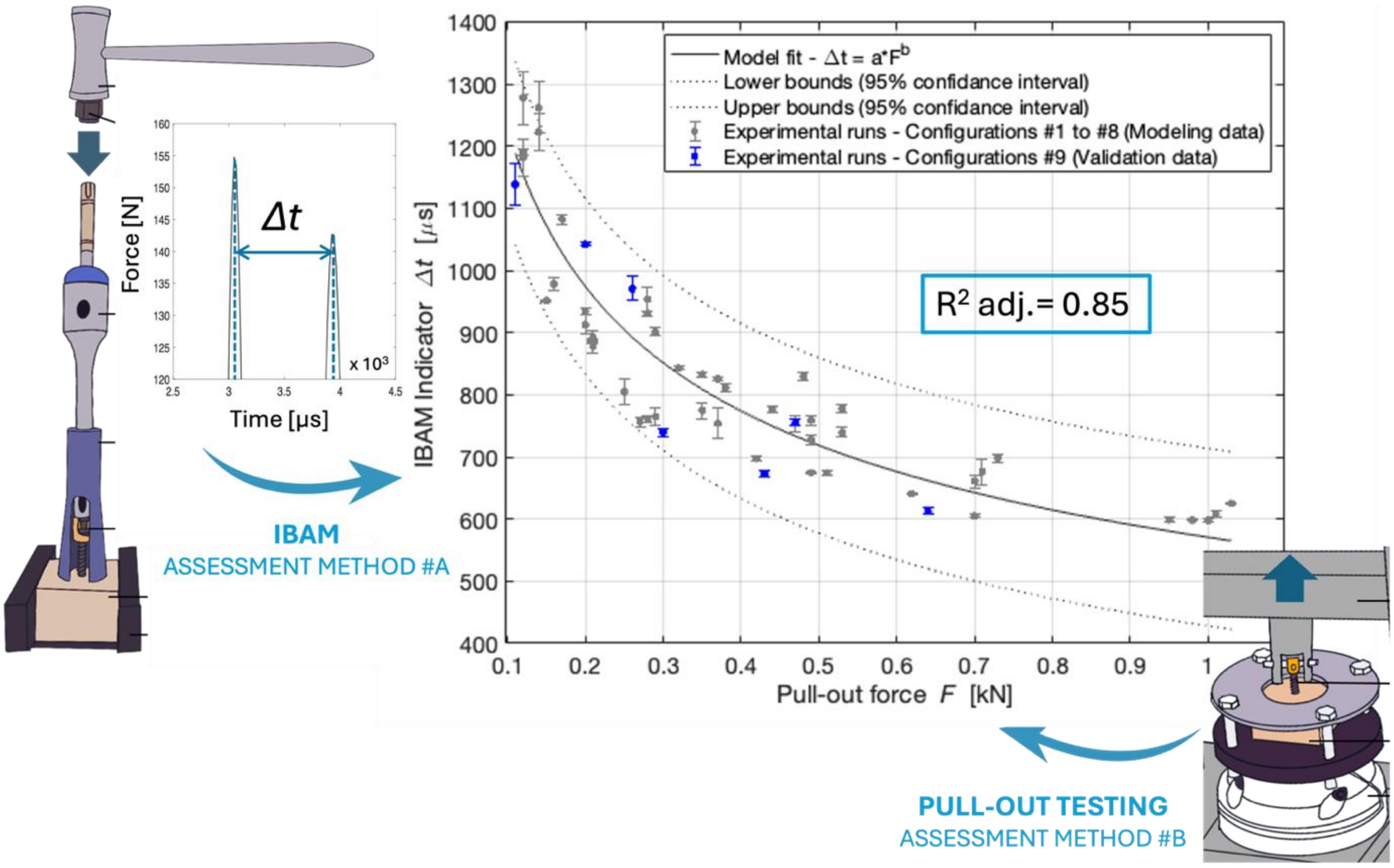

*Graphical abstract*

# I. Introduction

Among the available treatment options for spine disorders, spinal instrumentation remains the gold standard approach for correcting deformities, injuries, or degenerative conditions, due to its ability to restore spinal stability and alignment[1]. This technique relies on securing metallic rods onto pedicle screws (PS) implanted within the vertebral body to provide structural support to the spine[2].

The long-term success of this intervention strongly depends on the stability of the PS, which is determined by the evolution of the biomechanical properties of the bone-implant interface (BII) over time[3]. Immediately after insertion, the implant's initial fixation, known as primary stability, relies on the pre-stressed state of the surrounding vertebral bone tissues. Primary stability plays a crucial role in the bone regeneration phenomena (osteointegration) occurring around the PS during the healing process. It promotes an increase in the bone-implant contact (BIC) ratio which, ultimately, secures the permanent fixation of the implant, known as secondary stability.

As the screw is driven in, its threads engage and interlock with the bone tissues, inducing shear, tensile, and compressive stresses at the BII[4]. This biomechanical fixation results from a complex interplay between multiple variables, including patient-specific parameters (e.g. bone quality[5], vertebral anatomy[6]), the surgical technique (e.g. pilot hole preparation[7], trajectory definition[8]), and the implant design (e.g. screw diameter[9], thread distribution[10]). Achieving optimal primary stability therefore requires the surgeon to make a careful compromise between these variables. Excessive press-fitted state or oversized screw dimensions can over-constrain the surrounding tissues, potentially leading to local bone necrosis or fractures. Conversely, insufficient fixation may allow micromobilities at the BII, impairing bone regeneration and risking instrumentation failure due to PS aseptic loosening[11]. When insufficient fixation is anticipated (e.g. osteoporotic patients), surgeons may resort to cement augmentation. However, this approach introduces additional risks, including cement leakage, pulmonary embolism, adjacent compression in the vertebral disks, and infectious complications[12]. Furthermore, revision surgeries involving cemented screws are particularly complex and challenging.

Despite the importance of the PS primary stability, the current intraoperative method to assess its quality is restricted to the haptic feedback of the surgeon during and after the implant insertion. However, under experimental conditions, the assessment of PS stability can be performed using quantitative but destructive methods. Specifically, axial pull-out[13] or toggle tests[14] primarily assess the implant's mechanical resistance to loading, while push-based[15] and torque-based[16] measurements provide insight into the underlying bone properties that govern implant anchorage. Alternatively, *in silico* models[17,18] provide a fully non-invasive approach to predict PS stability, but their development necessitates a compromise between computational efficiency and predictive accuracy, and their translation into clinical practice remains limited by the scarcity of validation data. Moreover, several vibration-based techniques have been developed to address the need for quantitative intraoperative evaluation of the PS stability. In their early development, these methods were typically benchmarked against conventional mechanical tests. The first technique, known as modal analysis, consists of generating a vibration at the BII by striking the PS[19] or a customized system[20] with a hammer, or non-invasively, using non-contact approaches such as pulsed-laser-based techniques[21]. The resulting resonance frequency, depending on the biomechanical properties of the BII, was shown to be correlated with the PS pull-test force ($R^2 = 0.81$ for classical analysis[19] and $R^2 = 0.93$ for laser analysis[21]). Moreover, a medical device (Osstell®, Osstell AB, SWEDEN) originally developed for dental implant assessment has been repurposed for the PS[22]. While this system is designed to provide a stability score, it still requires the insertion of a SmartPeg inside the screw head. This additional component introduces variability in the scoring process. Factors such as misalignment, incomplete seating, or geometric constraints within the screw head may affect the accuracy and consistency of the measurements. Despite promising performances, the clinical adoption of these methods remains limited due to their invasiveness (requiring direct contact with the implant) and challenging implementation (logistical burden, personal protective equipment).

Interestingly, our group has developed a noninvasive Impact-Based Analysis Method (IBAM) relying on the use of an instrumented hammer to evaluate hip implants' primary stability. This hammer was equipped with a force sensor on its impact face, enabling to record the force signals generated during the implants' impaction. A significant correlation was found between the impact momentum and the implant pull-out force, with $R^2 = 0.69$ for the acetabular cup[23] and $R^2 = 0.81$ for the femoral stem[24]. Moreover, the time difference between the first two local maxima of the force signal, denoted *dt*, was shown to be a reliable indicator of bone-implant system stiffness[25]. IBAM has also been applied to different osteotomy procedures[26–29] and to soft tissues[30,31]. This study aims to investigate the feasibility of using IBAM to determine the primary stability of a PS inserted in bone mimicking phantoms. Since the PS is not impacted but screwed within bone tissue, we applied low-amplitude impacts onto the screwdriver once the implant has been inserted, and we analyzed the variation of *dt* under different stability configurations created in a controlled environment following an experimental design technique. As standard practice, the results obtained with our method were benchmarked against the PS axial pull-out force to validate its relevance.

# II. Materials and Methods

## 1. Experimental measurements

### a) Sample Preparation

Artificial bone samples (SawBones, Pacific Research, WA, USA) of dimensions 55 x 55 x 40 $mm^3$ were used with material densities varying according to the design of experiments (DoE) described in subsection II.2. Each sample was perforated centrally through its full width using a column drill operating at 460 rpm. Drill diameters were also determined according to the DoE for pilot hole diameters.

The same polyaxial self-tapped PS (7575540, Horizon® Legacy™ 5.5, Medtronic, IRELAND) was used for all experimental runs. The screw had a diameter of 5.5 mm and featured a uniform 3-mm thread pitch along a 40 mm shaft. To ensure accurate screw placement and minimize micromotions, each sample was securely clamped inside a vice (see Figure 1), while the screw was inserted to a depth defined by the DoE using the corresponding manual screwdriver (7880113, Horizon® Legacy™ 5.5). We hypothesize that the lateral pressure exerted by the vice does not affect the mechanical configuration, as sample length was defined to build an infinite domain (surface dimensions ~ 5x nominal PS diameter as recommended by ASTM F543-23[32]) and minimal efforts were applied during clamping.

The implant trajectory followed the axis of a customized insertion guide to control angulation and reduce experimental variability[33]. Depth insertion measurement error was estimated to be equal to 0.9 mm, based on repeated caliper measurements of the emerging screw shaft.

### b) Impact-Based Analysis Method (IBAM)

After the PS was screwed inside each bone phantom, maintained with the screwdriver handle (836-010, Medtronic, IRELAND) was removed, exposing a system composed of three parts (the screwdriver sleeve, the PS, and the bone sample) on which the IBAM was applied (see Figure 1A). The screwdriver sleeve was retained within the impact assembly to maintain a minimally invasive approach while providing optimal access to the impact zone.

A standard 250 g surgical hammer (32-6906-26, Zepf, GERMANY) equipped with a force sensor (208C04, PCB Piezoelectronics, NY, USA) in the center of its impaction face – the instrumented hammer (IH) – was used to manually perform axial impacts on the top of the screwdriver sleeve. Each impact force, defined as the maximum value of the force as a function of time, ranged between 100 N (to obtain a sufficient energy to excite the system) and 300 N (to preserve the integrity of the bone). The resulting time variation was recorded for a duration of 20 ms by a data acquisition module (NI-9234, National Instruments, TX, USA) with a sampling frequency of 51.2 kHz and transferred to a computer using a LabVIEW interface (National Instruments, TX, USA). The choice of the sampling frequency resulted from a compromise between signal transcription quality and data processing time. An example of a force signal $s(t)$ and time variation $dt$ are illustrated in Figure 1B.

For each experimental run, the IBAM consisted in averaging the response obtained for 5 consecutive impacts. To assess the reproducibility of the method, this process was repeated 4 times, resulting in a total of 20 impacts performed on the screwdriver sleeve. The IBAM indicator $\Delta t$ was calculated as the average of the 4 repetitions, following Equation 1:

$$\Delta t = \frac{1}{4}\sum_{j=1}^{4}\left(\frac{1}{5}\sum_{i=1}^{5} dt_i^j\right), \quad (1)$$

where $dt_i^j$ was determined for the *i-th* impact in the *j-th* repetition by the time difference between the two earliest local maxima of the force signal, as shown in Fig. 1B. A flow chart describing the procedure used to compute the indicator $\Delta t$ is provided in Appendix C. The standard deviation of the averages obtained for the 4 repetitions was also determined for each experimental run to assess the reproducibility of the IBAM.

*Figure 1: A) Schematic representation of the experimental setup composed of the impacted system on which the IBAM was performed with low-amplitude axial impact using an instrumented hammer. B) Example of a force signal recorded during an axial impact on the surgical screwdriver for configuration #6 and repetition #2.*

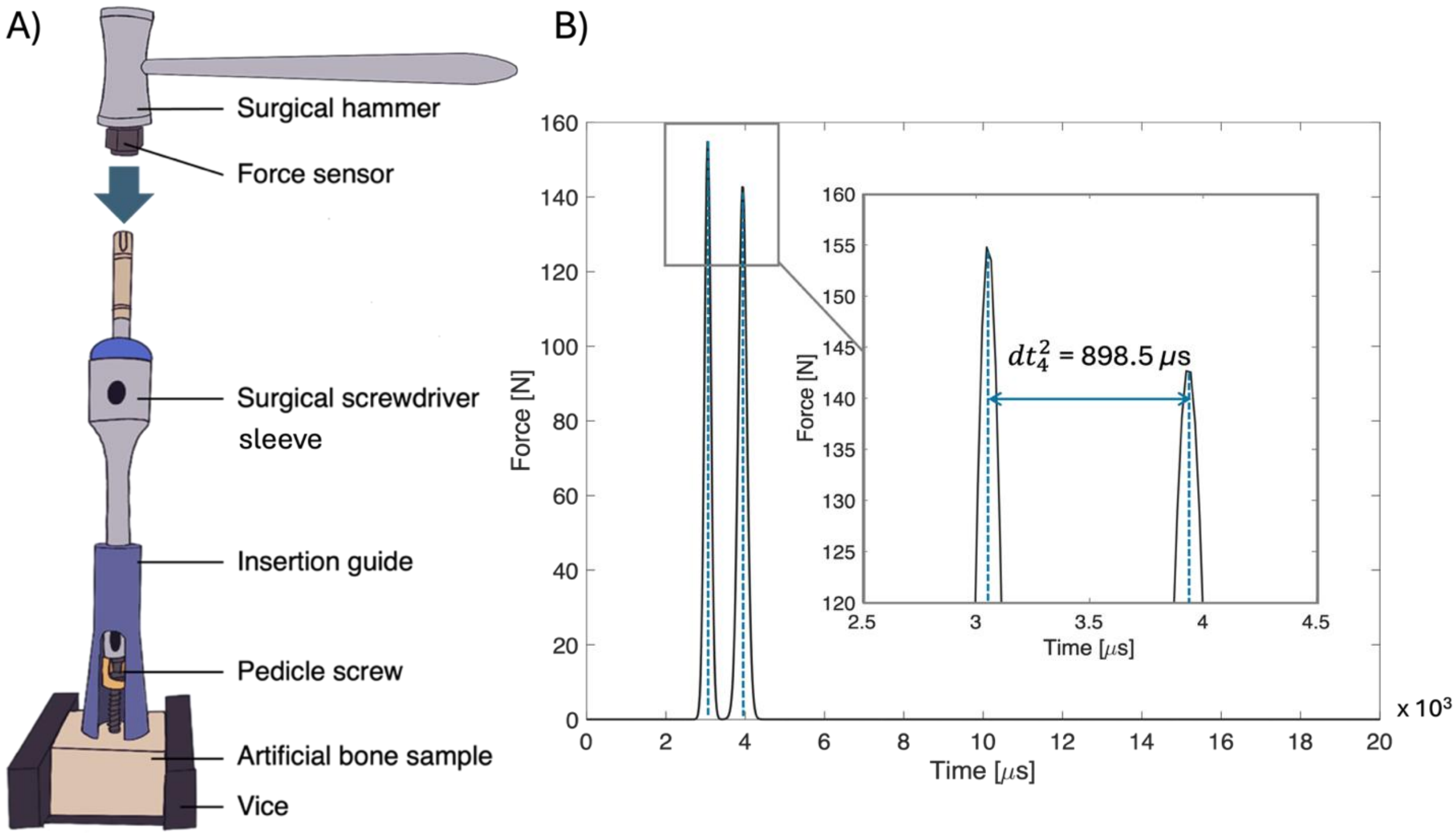

c) Pull-Out Testing

After performing IBAM, the screwdriver sleeve was removed from the PS head and the sample was positioned into a universal testing machine (DY25, Adamel Lhomargy, Roissy en Brie, FRANCE) to perform pull-out testing, as illustrated in Figure 2. The test fixture was adjusted to align the implant's longitudinal axis with the pull-out direction. The screw was connected to the machine actuator by fixing a rigid stem into the screw head. A quasi-static tensile load was applied with a constant displacement rate of 2 mm/min until the screw was fully extracted from the bone sample. The maximum force recorded during this procedure was defined as the pull-out force $F$ of the PS and corresponds to the maximum force required to extract the PS. This force was used as a response variable inside the DoE.

*Figure 1: Schematic representation of the pull-out testing setup*

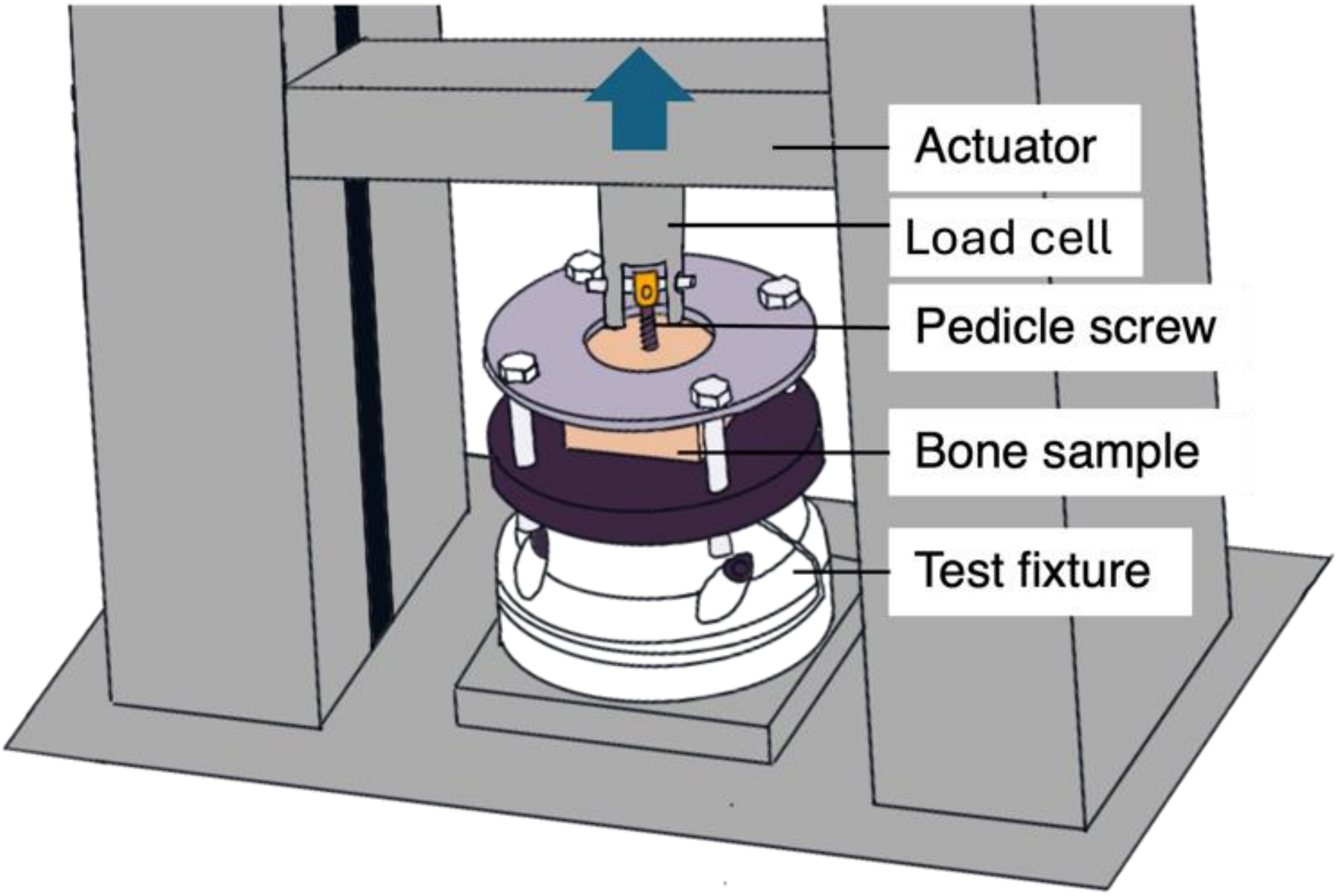

## 2. Multi-Response Design of Experiment (DoE)

### a) Full Factorial Design

We considered a multi-response design of experiment (DoE) by employing a central composite design (CCD)[34] to compare the performances of IBAM and of axial pull-out testing. This statistical approach enabled the modeling of the main and interaction effects of 3 experimental factors on the response variable associated with each technique. Specifically, main effects represent the individual impact of each experimental factor on the response variable, while interaction effects describe how the combination of multiple factors jointly influences the response. Statistical significance and reliability of these effects were assessed with 95% confidence intervals ($\alpha = 0.05$) using the method described by Box and Wilson[34].

The CCD factorial space defined 8 configurations combining high and low factor levels of the experimental factors. Each configuration was repeated six times to ensure robustness, resulting in 48 experimental runs. For each experimental run, both response variables were measured as explained in section II.1.

b) Experimental Factors Choice

Artificial bone density ($\rho$), insertion depth ($D$), and pilot hole diameter ($\phi$) were selected as experimental factors due to their well-established effect on the PS primary stability[5,7,14]. The analysis was intentionally restricted to a limited number of 3 factors to preserve a simple and robust statistical framework. In Table 1, High-level factors (+) positively impact the PS primary stability, while Low-level factors (-) disadvantage its initial biomechanical anchorage. Their respective numerical values were chosen to balance clinical relevance with experimental feasibility. The artificial bone density variability captures a full spectrum of bone conditions from healthy (15 PCF) to osteoporotic (10 PCF) vertebral tissues[13,35]. The minimum insertion depth (6 screw turns) was chosen to ensure screw immobility, while the maximum depth (11 screw turns) was set to prevent the screw head from contacting the artificial bone, as its influence on the implant's primary stability remains poorly understood[36]. Finally, the pilot hole diameter range (from 60 to 80% of the PS nominal diameter) was selected to reduce experimental variability as the interaction between the screw thread and the bone tends to produce inconsistent and difficult-to-replicate outcomes. A smaller pilot hole leaves more material in contact with the threads, increasing friction and variability, while a larger pilot hole may compromise implant immobility due to insufficient bone engagement.

| | **LEVEL** | | |
|---|---|---|---|
| **FACTOR** | **High (+)** | **Medium (0)** | **Low (-)** |
| **Artificial bone density $\rho$ [PCF]** | 15 | 12 | 10 |
| **Insertion depth $D$ [Screw Turn]** | 6 | 8 | 11 |
| **Pilot hole diameter $\phi$ [% of the PS nominal diameter]** | 60 | 70 | 80 |

*Table 1 : Experimental factor levels*

An additional configuration was included at the center of the CCD factorial space with all factors set to medium level. This configuration was also repeated 6 times, giving a total of 54 experimental runs. A Lack-of-Fit test, conducted through ANOVA using MATLAB (Mathworks, MA, USA) assessed the goodness of fit of the model and confirmed the overall significance of the main and interaction effects.

# III. Results

The experimental matrix (see Table 2) displays the mean and standard deviation values of the response variables for each configuration tested, with raw data shown in Appendices A and B. These values were extracted for each of the eight CCD modeling configurations (#1 to #8) as well as for the additional validation configuration (#9). The configuration yielding the highest PS primary stability was #1, characterized by all factors set at their High levels. In this configuration, the response variable $F$ reaches its maximum value ($F = 0.10 \pm 0.03$ kN) while the IBAM indicator is at its minimum ($\Delta t = 637.8 \pm 79.1$ µs). Conversely, configuration #8, where all factors are set at their Low levels, resulted in the lowest pull-out force ($F = 0.13 \pm 0.01$ kN) and highest $\Delta t$ value ($\Delta t = 1212.1 \pm 52.5$ µs).

| | FACTOR LEVEL COMBINATION | | | RESPONSE VARIABLES | | | |
|---|---|---|---|---|---|---|---|
| Configuration # | Artificial bone density ($\rho$) | Insertion depth ($D$) | Pilot hole diameter ($\phi$) | $\Delta t$ [µs] | | $F$ [kN] | |
| | | | | Mean | STD | Mean | STD |
| **1** | + | + | + | 637.8 | 79.1 | 1.00 | 0.03 |
| **2** | - | + | + | 800.5 | 30.1 | 0.35 | 0.05 |
| **3** | + | - | + | 706.0 | 43.3 | 0.49 | 0.05 |
| **4** | - | - | + | 979.6 | 71.8 | 0.18 | 0.02 |
| **5** | + | + | - | 649.0 | 36.5 | 0.68 | 0.04 |
| **6** | - | + | - | 918.1 | 36.4 | 0.24 | 0.04 |
| **7** | + | - | - | 799.3 | 74.0 | 0.29 | 0.08 |
| **8** | - | - | - | 1212.1 | 52.5 | 0.13 | 0.01 |
| **9** | 0 | 0 | 0 | 767.1 | 34.4 | 0.47 | 0.02 |

*Table 2: Mean and standard deviation (STD) of the response variables, $\Delta t$ and $F$, across the repetitions of each configuration, combining High (+), Medium (0) and Low (-) levels of Bone density ($\rho$), Insertion depth ($D$) and Pilot hole diameter ($\phi$).*

The DoE and ANOVA analyses confirm a significant influence ($p < 0.05$) of all 3 factors ($\rho$, $D$ and $\phi$) on both response variables, as shown in Figure 3, which presents a bar graph ranking the main and interaction effects (Pareto chart) of the experimental factors. While each main effect contributes similarly in magnitude to both response variables, the direction of the effect is reversed between them. Additionally, the main effects exhibit a greater influence (~78% of contribution) than the interaction effects on both measures. Regarding interaction effects, all are found to significantly influence the pull-out force $F$, whereas the interaction between $D$ and $\phi$ has no significant impact on $\Delta t$. However, the interaction between $D$ and $\phi$ was found to be the least influential on $F$ (3.7% of contribution), which supports the observed consistency between the two response variables. Finally, the lack-of-fit test indicates a strong model fit, with adjusted $R^2$ values of 0.98 and 0.90 for the pull-out force and IBAM indicator, respectively.

*Figure 2: Pareto chart ranking the main and interaction effects on the response variables, F (left) and Δt (right) according to their p-value determined with ANOVA. The experimental factors considered were Artificial bone density (ρ), Insertion depth (D) and Pilot hole diameter (ϕ). The contribution of each significant effect on the response variables is indicated as a percentage*

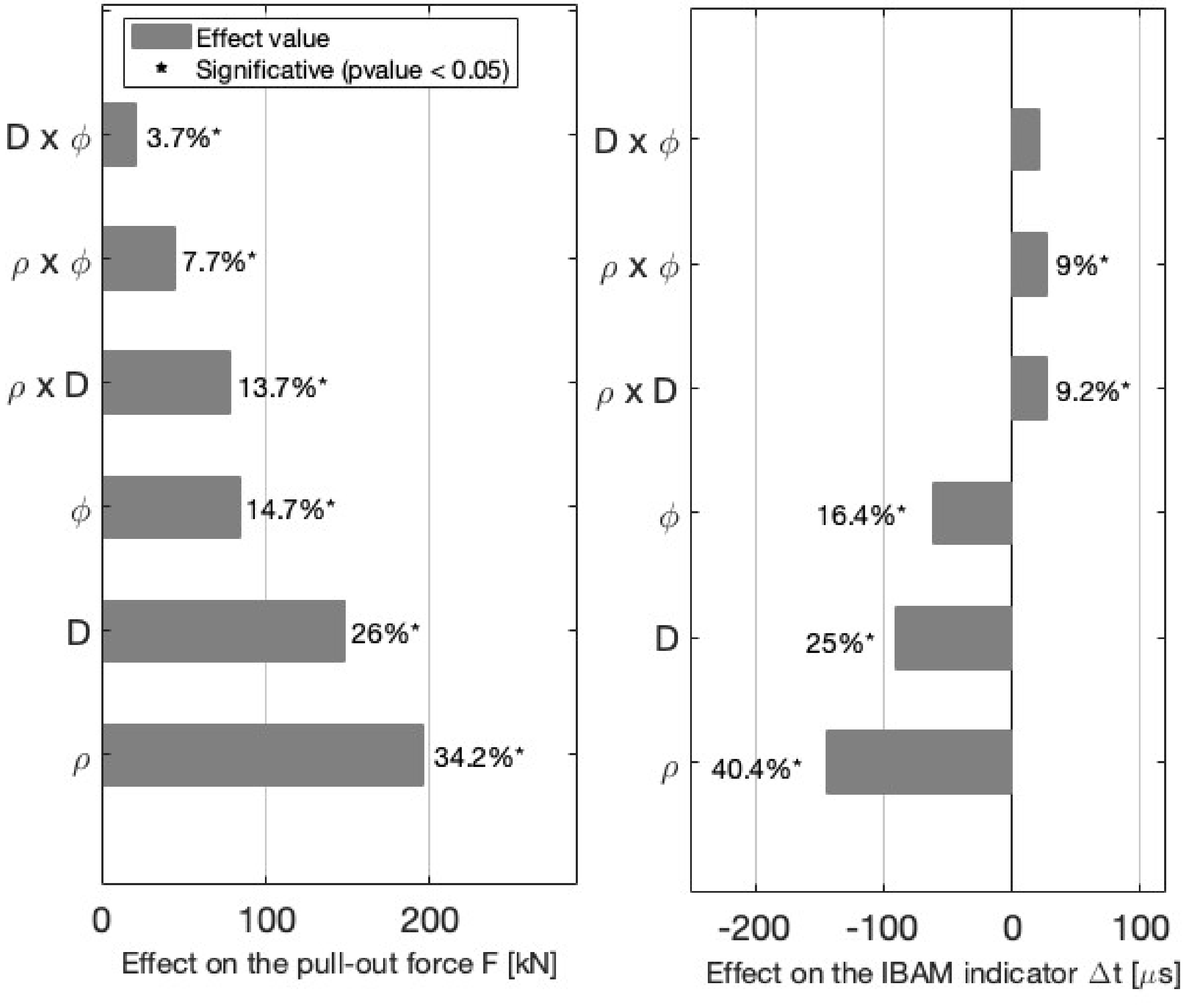

Figure 4 shows that $\Delta t$ decreases with the increasing pull-out force $F$ according to a power-law relationship ($\Delta t = a*F^b$), with a positive scaling factor $a = 0.0006$ and negative exponent $b = - 0.3662$, resulting in an adjusted $R^2$ of 0.85.

*Figure 4: Variation of the IBAM indicator Δt as a function of pull-out force F. Error bars represent the measurement variability observed during the 20 impacts on the screwdriver*

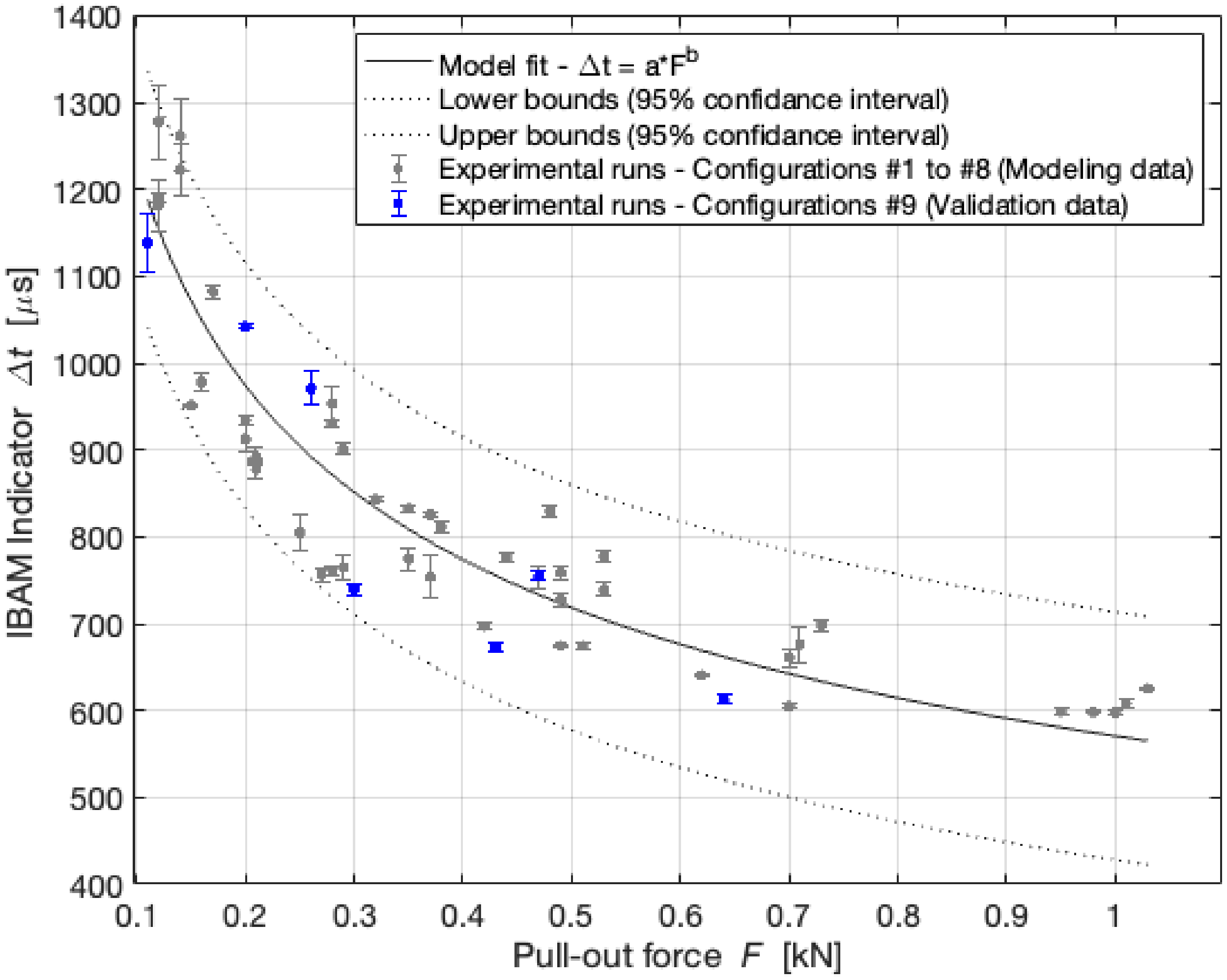

The significant interaction effects common to both response variables are presented in Figure 5, illustrating how bone density $\rho$ modulates the effects of $D$ and $\phi$ on both response variables. For pull-out force $F$, Figure 5 shows steeper slopes (+3.2% variation for $\phi$ and +3% variation for $D$) with respect to both decreasing $\phi$ and increasing $D$ at higher density levels, indicating that the associated stability gain is more readily measurable in high-density bone. Conversely, for IBAM, steeper slopes (+10.5% variation for $\phi$ and +2.5% variation for $D$) are observed at lower density levels, suggesting that stability gain is more measurable at low density when assessed using IBAM.

*Figure 5: Significant interaction effects: density (ρ) and pilot hole diameter (ϕ) on the left (D constant), density (ρ) and Insertion depth (D) on the right (ϕ constant)*

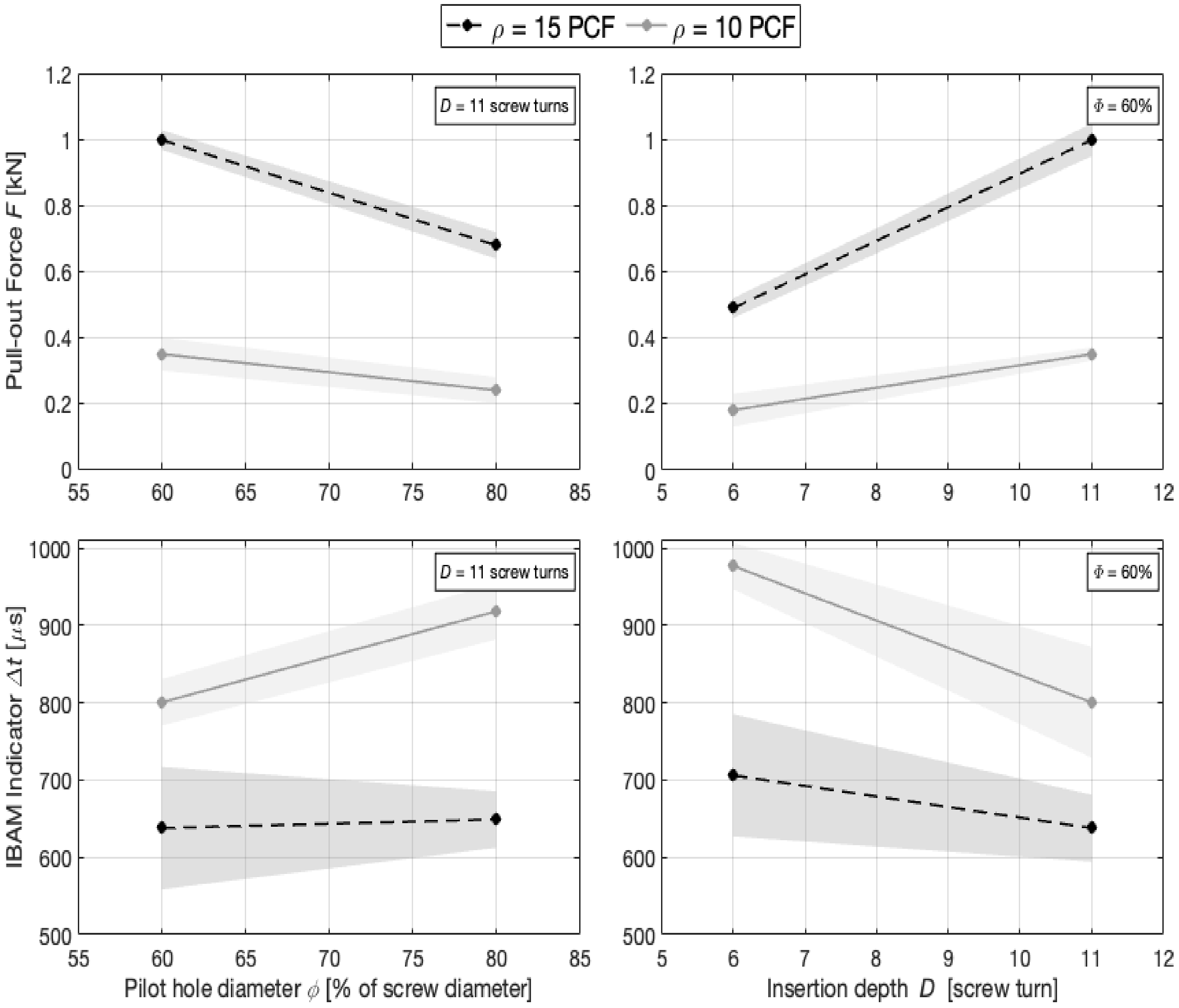

# IV. Discussion

This proof-of-concept study lays the groundwork for a novel approach to quantify PS primary stability in a non-invasive manner using an instrumented hammer. Although direct clinical applicability fell beyond the scope of this work, its results highlight promising avenues for further development of the method. By delivering low-amplitude impacts to the surgical construct, the IBAM may offer a cost-effective technique to assist surgeons in making informed intraoperative decisions, ultimately contributing to the improvement of spinal instrumentation surgery. A comparative approach between IBAM and axial pull-out testing was intended to strengthen our rationale regarding IBAM's potential to capture relevant properties of the bone-implant interface. This choice of comparative method was primarily motivated by the current standards[32] for PS homologation relying on an axial pull-out test performed in artificial bone-phantom as the reference method for clinical validation. Specifically, IBAM was benchmarked against axial pull-out testing for a PS inserted inside artificial bone samples with varying stability conditions controlled *via* a DoE framework. This structured and statistically grounded approach allowed for a comprehensive comparison of factorial effects across both assessment techniques.

The foundational concept behind IBAM was initially to monitor the impaction process of hip implants, ensuring safer implantation during arthroplasty surgery. Notably, the implant's primary stability was shown to be related to the impact momentum extracted from the impact force signal ($R^2$ = 0.69 for the acetabular cup[23] and $R^2$ = 0.81 for the femoral stem[24]). In this context, the time-variation $dt$ between the two earliest maxima of the force signal recorded for each impact evolved throughout implant insertion, allowing the determination of its optimally seated position. An analytical model[25] of this configuration demonstrated that $dt$ could be interpreted as the natural period of the system when its dynamics is approximated by a one-dimensional damped oscillator. Under this assumption, $dt$ decreases with the stiffness of the impacted system, which was confirmed by finite element studies[37,38]. In the present work, the static nature of the PS within the vertebra enabled the adaptation of the original concept for a non-invasive and non-destructive impaction process, allowing the extraction of valuable information from the corresponding force signals. Notably, this study is the first of our group to consider $dt$ as an indicator for implant stability assessment.

The final IBAM indicator $\Delta t$, calculated using Equation (1), follows the same variations as $dt$. Indicator $\Delta t$ showed an opposite dependence on the experimental parameters (Artificial bone density, insertion depth, and pilot hole diameter) compared to the pull-out force $F$ (see Figure 3). While the pull-out force $F$ increases with PS primary stability, $\Delta t$ decreases, consistently with the analytical[25] and previous experimental observations [23,24,39–41]. Here, the system stiffness is enhanced by increased bone density ($\rho$) and insertion depth ($D$), along with reduced pilot hole diameter ($\phi$), promoting a strong bone-implant ratio (BIC) at the BII[3]. While $F$ and $\Delta t$ provide complementary insights, their relationship is non-linear (see Figure 4), suggesting that each response variable captures distinct aspects of the PS primary stability. The nuance between the two measures becomes apparent when examining the interaction effects on both response variables (see Figure 5) as they exhibit opposite sensitivities to bone density level. While the pull-out force shows greater responsiveness to changes in $D$ and $\phi$ at higher bone density, IBAM demonstrates a stronger responsiveness at lower bone density. Overall, the significance of interaction effects suggests that the effects of pilot hole diameter and screw insertion are density dependent. From a clinical perspective, these findings reinforce current practice, highlighting the importance of considering bone density when planning the therapeutic strategy and tailoring the surgical protocol. Moreover, it would be interesting to compare the results obtained with IBAM and other conventional mechanical tests, such as torque measurement[16] or toggling[42], to further investigate the degree of similarity among those assessment methods. Specifically, advanced stages of IBAM's development will require a more comprehensive benchmarking strategy to fully characterize its robustness.

Furthermore, the $\Delta t$ extraction protocol described by Equation 1 provides insights into the measurement variability of our method. Considering the data distribution shown in Figure 4, the highest measurement variations of $\Delta t$ were obtained for configurations with the weakest pull-out forces, which may be attributed to the difficulty of applying reproducible impacts to an unstable configuration. Interestingly, despite these configuration-dependent variations, the maximal relative uncertainty of $\Delta t$ remains lower to 1%, reflecting a high level of precision. Therefore, this variability does not indicate an intrinsic limitation of the IBAM, but rather reflects the current experimental implementation, and particularly the limited ergonomics of the proof-of-concept setup used in this study. Note that the number of $dt$ values currently considered in the calculation of $\Delta t$ resulted from a compromise between achieving sufficient reproducibility, repeatability, and measurement efficiency. Future studies working toward clinical application should optimize the measurement protocol and instrumentation to improve ergonomics and efficiency while maintaining measurement precision.

Although the current instrumented hammer used to extract $\Delta t$ cannot be directly transferrable to the clinic, the underlying principle of IBAM exhibits strong potential to become an intraoperative method to measure PS primary stability. Its transferability to clinical practice represents a technical and engineering challenge, which lies beyond the scope of the present work and is left for future investigations. For example, a miniaturized punch could be embedded into a screwdriver-like tool, commonly used during PS insertion, enabling intraoperative measurements without disrupting standard procedures. Such an embedded design would not introduce any additional surgical step but rather augment an existing instruments with sensing capabilities, thereby transforming it into a decision support tool. In situations where the PS primary stability is insufficient, such tool may allow the surgeon to adapt the therapeutic strategy in real-time, for example by selecting a larger implant or by using cement augmentation. Additionally, IBAM's application would be greatly beneficial as it is a non-invasive, non-ionizing, low-cost method that does not require specialized operator training. Nevertheless, the method must be first validated in appropriate animal models, in accordance with standard preclinical methodology.

There are several limitations to our study. The complexity of the interactions between the PS and the bone sample is not fully accounted for by the current experimental design despite the accuracy of the model fit ($R^2_F = 0.98$ and $R^2_{\Delta t} = 0.90$). Consequently, the model derived from factorial effect estimation should be interpreted with caution and should not be used for predictive purposes. Moreover, while this study considers polyurethane foam samples designed to mimic the dynamical behavior of the bone, the current experimental configuration does not reproduce the biological and structural complexity of vertebral tissues. From a bone-centered perspective, the model fails to capture the interactions between cortical and trabecular structures that critically influence the PS primary stability, such as the pedicle[6]. When considering the spine as a mechanical system, our approach could be influenced by the structural arrangement and patient-specific variability of the intervertebral discs and adjacent vertebrae. In addition, the oscillatory behavior of the construct may be damped by the soft tissues and spine mobility. Note that we have shown in a previous study using IBAM for assessment of the acetabular cup stability[43] that the surrounding soft tissue did not affect the results, which needs to be confirmed in the context of the present study. Consequently, further development of the IBAM should include feasibility assessment within a full-spine or spinal segment models.

# V. Conclusions

This study introduces an Impact-based Analysis Method (IBAM) to evaluate the primary stability of pedicle screws (PS) by performing low-amplitude impacts on a configuration rigidly bounded with the implant. We considered variations of the PS initial fixation using experimental design techniques to compare the time variation between the two earliest local maxima of the corresponding force signals and the PS pull-out force. While these two measurements capture different aspects of primary stability, strong similarities were observed. Further studies are needed to validate its performance and applicability across various clinical scenarios. Nevertheless, this work highlights the opportunity to develop IBAM as a non-invasive and cost-effective technique to assist surgeons in making informed intraoperative decisions, ultimately contributing to the improvement of spinal instrumentation procedures. Moreover, IBAM could be extended to evaluate the stability of screws used in other orthopedic applications, such as fracture reduction interventions in traumatology, to assess the quality of its stabilization.

# VI. Acknowledgements

This work has received funding from the project OrthAncil (ANR-21-CE19-0035-03), from the project OrthoMat (ANR-21-CE17-0004), from the project ModyBe (ANR-23-CE45-0011-02), from the project IBAM (ANR-24-LCV2-0009-01), from the project MaxilOsteo (ANR-24-CE17-2236-01), from the European Research Council (ERC) under the European Union’s Horizon 2020 research and innovation program (Grant Agreement N°101062467, project ERC Proof of concept 2022 Impactor) and from the Natural Science and Engineering Research Council of Canada.

# Appendices

| Configuration # | FACTOR LEVEL COMBINATION | | | *INDICATOR* Δ*t* [µs] | | | | | |
|---|---|---|---|---|---|---|---|---|---|
| | Artificial bone density ($\rho$) | Insertion depth ($D$) | Pilot hole diameter ($\phi$) | Repetition | | | | | |
| | | | | #1 | #2 | #3 | #4 | #5 | #6 |
| **1** | + | + | + | 625.1 | 625.3 | 608.1 | 598.2 | 598.9 | 598.5 |
| **2** | - | + | + | 804.9 | 774.5 | 754.2 | 825.9 | 832.7 | 810.1 |
| **3** | + | - | + | 697.3 | 739.7 | 777.3 | 674.4 | 674.2 | 673.2 |
| **4** | - | - | + | 934.0 | 978.3 | 1082.6 | 951.6 | 888.4 | 1042.4 |
| **5** | + | + | - | 640.6 | 676.5 | 661.1 | 698.3 | 605.1 | 612.5 |
| **6** | - | + | - | 912.6 | 892.8 | 900.9 | 877.8 | 953.5 | 970.8 |
| **7** | + | - | - | 931.0 | 756.3 | 843.3 | 761.2 | 764.8 | 739.4 |
| **8** | - | - | - | 126.2 | 127.8 | 118.1 | 119.0 | 122.3 | 113.9 |
| **9** | 0 | 0 | 0 | 776.1 | 727.8 | 829.8 | 759.3 | 754.3 | 755.5 |

*Appendix A : Response variable* Δ*t across the repetitions of each configuration, combining High (+), Medium (0) and Low (-) levels of Bone density ($\rho$), Insertion depth (D) and Pilot hole diameter ($\phi$).*

| Configuration # | FACTOR LEVEL COMBINATION | | | PULL-OUT FORCE F [kN] | | | | | |
|---|---|---|---|---|---|---|---|---|---|
| | Artificial bone density ($\rho$) | Insertion depth ($D$) | Pilot hole diameter ($\phi$) | Repetition | | | | | |
| | | | | #1 | #2 | #3 | #4 | #5 | #6 |
| **1** | + | + | + | 1.03 | 1.03 | 1.01 | 1.00 | 0.95 | 0.98 |
| **2** | - | + | + | 0.25 | 0.35 | 0.37 | 0.37 | 0.35 | 0.38 |
| **3** | + | - | + | 0.42 | 0.53 | 0.53 | 0.51 | 0.49 | 0.43 |
| **4** | - | - | + | 0.20 | 0.16 | 0.17 | 0.15 | 0.21 | 0.20 |
| **5** | + | + | - | 0.62 | 0.71 | 0.70 | 0.73 | 0.70 | 0.64 |
| **6** | - | + | - | 0.20 | 0.21 | 0.29 | 0.21 | 0.28 | 0.26 |
| **7** | + | - | - | 0.28 | 0.27 | 0.32 | 0.28 | 0.29 | 0.30 |
| **8** | - | - | - | 0.14 | 0.12 | 0.12 | 0.12 | 0.14 | 0.11 |
| **9** | 0 | 0 | 0 | 0.44 | 0.49 | 0.48 | 0.49 | 0.47 | 0.47 |

*Appendix B: Response variable F across the repetitions of each configuration, combining High (+), Medium (0) and Low (-) levels of Bone density ($\rho$), Insertion depth (D) and Pilot hole diameter ($\phi$).*

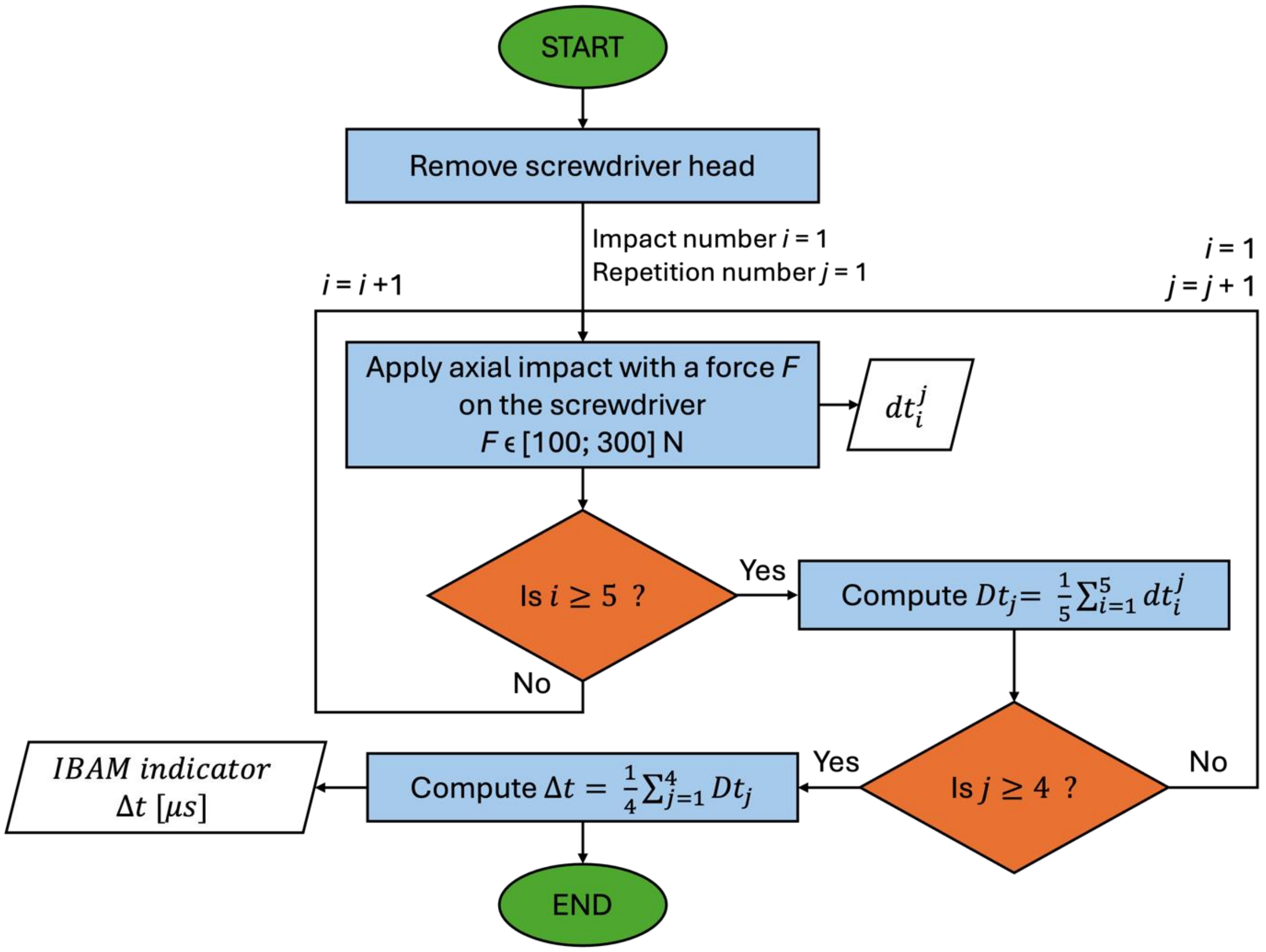

*Appendix C: Flow Chart of the IBAM procedure*